\documentclass[
]{ceurart}

\usepackage{listings}
\usepackage{tabularx}
\usepackage{booktabs}
\usepackage{multirow}   
\usepackage{siunitx}
\usepackage{placeins}
\begin{document}

\copyrightyear{2025}
\copyrightclause{Copyright for this paper by its authors.
  Use permitted under Creative Commons License Attribution 4.0
  International (CC BY 4.0).}

\conference{GeCoIn 2025: Generative Code Intelligence Workshop, co-located with the 28th European Conference on Artificial Intelligence (ECAI-2025), October 26, 2025 --- Bologna, Italy}

\title{DeepCodeSeek: Real-Time API Retrieval for Context-Aware Code Generation}

\author[1]{Esakkivel Esakkiraja}[%
email=esakkivel.esakkiraja@servicenow.com,
]

\author[1]{Denis Akhiyarov}[%
email=denis.akhiyarov@servicenow.com,
]
\cormark[1]

\author[1]{Aditya Shanmugham}[%
email=aditya.shanmugham@servicenow.com,
]

\author[1]{Chitra Ganapathy}[%
email=chitra.ganapathy@servicenow.com,
]

\address[1]{ServiceNow, Inc.}

\cortext[1]{Corresponding author.}

\begin{abstract}
   Current search techniques are limited to standard RAG query-document applications. In this paper, we propose a novel technique to expand the code and index for predicting the required APIs, directly enabling high-quality, end-to-end code generation for auto-completion and agentic AI applications. We address the problem of API leaks in current code-to-code benchmark datasets by introducing a new dataset built from real-world ServiceNow Script Includes that capture the challenge of unclear API usage intent in the code. Our evaluation metrics show that this method achieves 87.86\% top-40 retrieval accuracy, allowing the critical context with APIs needed for successful downstream code generation. To enable real-time predictions, we develop a comprehensive post-training pipeline that optimizes a compact 0.6B reranker through synthetic dataset generation, supervised fine-tuning, and reinforcement learning. This approach enables our compact reranker to outperform a much larger 8B model while maintaining 2.5x reduced latency, effectively addressing the nuances of enterprise-specific code without the computational overhead of larger models.
\end{abstract}


\maketitle

\section{Introduction}
\label{sec:intro}

Large Language Models (LLMs) have become integral to modern developer workflows through AI-assisted code completion. In specialized enterprise environments like ServiceNow, model effectiveness depends heavily on context quality, particularly for custom APIs called Script Includes. Script Includes in ServiceNow are reusable JavaScript components that serve as a centralized repository for storing functions and classes, enabling developers to encapsulate complex business logic \cite{snow2025si}.

This paper addresses the critical challenge of context retrieval for LLM-powered code generation in ServiceNow's code completion and Build Agent tasks.

The core problem is accurately retrieving relevant Script Includes from partial developer code without explicit queries. Traditional methods like keyword search or basic vector search fail to capture nuanced developer intent and lack awareness of complex hierarchical relationships across the ServiceNow platform. General-purpose LLMs also lack domain-specific knowledge, making high-quality retrieval essential for reusing instance-specific Script Includes.

We propose DeepCodeSeek, a multi-stage retrieval pipeline that maximizes context relevance for LLMs. Our main contributions are: (1) a search pipeline using platform metadata and advanced IR techniques to significantly improve retrieval accuracy over baselines; (2) a comprehensive post-training pipeline optimizing compact reranker models through synthetic dataset generation, supervised fine-tuning, and reinforcement learning; and (3) empirical validation showing our optimized 0.6B reranker surpasses 8B models while maintaining significantly reduced latency for real-time applications.

The rest of this paper is organized as follows: Section~\ref{sec:proposed_method} details our multi-stage retrieval method, Section~\ref{sec:dataset_and_index} describes dataset construction and indexing, Section~\ref{sec:experimental_setup} presents experimental setup and evaluation methodology, Section~\ref{sec:experiments} shows main results and ablation studies, and Section~\ref{sec:post_training} details our post-training pipeline for optimizing compact reranker models.

\section{Related Work}
\label{sec:related_work}

Our work builds on recent advances in neural code retrieval, retrieval-augmented generation (RAG), structural code analysis, and search refinement techniques, adapting them to a large-scale enterprise environment.

\subsection{Neural Code Retrieval}
Code search has evolved from keyword-based methods like BM25 \cite{robertson2009probabilistic} to dense retrieval models such as CodeBERT \cite{feng2020codebert}, which embed queries and code into a shared semantic space. Yet recent evaluations (e.g., CoIR \cite{li2024coir}) show that general-purpose dense retrievers degrade in domains different from what they were trained on, mainly because their pretraining rarely covers such niche knowledge. To control for this effect, we adopt BM25 as a strong, domain-agnostic baseline that remains competitive under out-of-domain conditions. Our work then targets the missing piece: a domain-aware retrieval pipeline tailored to ServiceNow Script Includes.

\subsection{RAG with Filtering and Query Enhancement}
Retrieval-Augmented Generation (RAG) \cite{lewis2020retrieval} improves LLM outputs by dynamically providing relevant context, now common in coding assistants \cite{li2025conan}. A key challenge is ensuring retrieved context relevance, which can be addressed by leveraging code structure \cite{allamanis2018survey} and query enhancement techniques. Inspired by Hypothetical Document Embeddings \cite{gao2022precise}, LLMs can generate complete hypothetical code snippets from partial code, creating richer queries.

While many systems build Code Knowledge Graphs from source code \cite{liang2022kg4py, liu2024codexgraph} to enable filtering, our RAG pipeline constructs a Knowledge Graph from ServiceNow platform metadata for scope-level filtering. This constrains the search space and enables efficient retrieval of relevant Script Includes within our enterprise-specific context.

\subsection{Reranking}
Following retrieval, a cross-encoder or long-context LLM reranker \cite{sun2023chatgpt} can re-order top candidates, ensuring the most relevant results are prioritized for the final generation step. Recent reinforcement-learning approaches explicitly inject reasoning steps to boost reranking quality: REARANK \cite{zhang2025rearank} introduces a list-wise reasoning reranking agent, while SWE-RL \cite{wei2025swerl} shows that RL on large-scale software-evolution data substantially improves LLM reasoning for code-centric tasks. Our reranker adopts a similar RL fine-tuning strategy but is trained on enterprise-specific Script Include pairs, enabling higher precision in our domain.

\section{Proposed Method}
\label{sec:proposed_method}

\begin{figure}[t]
  \centering
  \includegraphics[width=\linewidth]{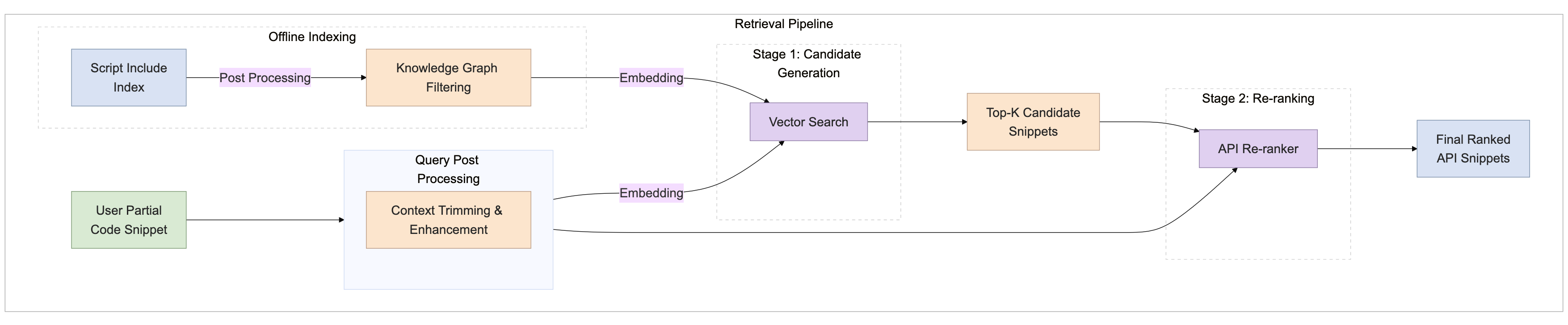}
  \caption{\textbf{Overview of the proposed multi-stage retrieval pipeline.}}
  \label{fig:model_teaser}
\end{figure}

Our approach is a multi-stage retrieval pipeline designed to provide highly relevant Script Includes for code completion. The pipeline begins with setting a baseline and incorporates several techniques to progressively refine the search space and improve accuracy. The overall architecture is depicted in Figure~\ref{fig:model_teaser}.

The core components of our method are as follows:

\begin{itemize}
    \item \textbf{Knowledge Graph for Search Space Reduction:} We leverage a Knowledge Graph (KG) constructed from platform metadata to prune the search space. This pre-filtering step significantly narrows the field of potential candidates before the main retrieval stage.

    \item \textbf{Enriched Indexing:} Rather than indexing the raw code, we created a structured index. All methods belonging to a single Script Include are grouped under their parent namespace. This index is further enriched with SI code metadata and their corresponding structured JSDoc, including API usage example. This organization helps the embedding model better distinguish between different functionalities and reduces ambiguity during retrieval.

    \item \textbf{LLM-Powered Code Expansion:} Developer's partial code often lacks sufficient context for effective retrieval. To address this, we experimented using a Large Language Model (LLM) at runtime to generate more descriptive and effective queries. By analyzing the partial code, the LLM can infer the developer's intent and produce a more complete code expansion, which in turn leads to more accurate results from the embedding model.

    \item \textbf{Reranking:} The initial retrieval stage may return the correct Script Include but not necessarily at the top of the list (e.g., within the top-5 results). For effective code generation, the downstream LLM needs a small, highly relevant set of options. Therefore, we employ a reranking stage using a cross-encoder or LLM reranker to improve the position of the most relevant candidates, aiming to move them from higher K values to lower K values (e.g., top-40 into the top-5). This ensures better performance, as it is easier for the code generation model to process fewer, higher-quality context options.
    
    \item \textbf{Post-training optimization:} We develop a comprehensive training pipeline that optimizes compact reranker models through synthetic dataset generation, supervised fine-tuning, and reinforcement learning, enabling smaller models to achieve performance comparable to much larger models while maintaining significantly reduced latency.
\end{itemize}

This multi-stage process, combining a knowledge-informed search space, enriched indexing, advanced query generation, and re-ranking, forms a robust pipeline that significantly outperforms vanilla retrieval methods(\cite{robertson2009probabilistic}) for code generation tasks.

\section{Dataset and Index Construction}
\label{sec:dataset_and_index}

\subsection{Dataset Construction}

We constructed a custom evaluation dataset from real-world ServiceNow development scenarios to capture the challenge of API retrieval from partial code. Our dataset consists of 850 code completion scenarios, each containing a partial JavaScript code snippet and the corresponding ground truth Script Include that should be used to complete the code.

To explain the terms used in our dataset:
\begin{itemize}
    \item \textbf{code\_middle}: Autocompletion span where the target API is invoked.
    \item \textbf{code\_before / code\_after}: Prefix and suffix around \texttt{code\_middle}; the prefix omits the target Script Include so retrieval must rely on context, while the suffix adds extra lines without exposing the API.
\end{itemize}

\begin{figure}[h]
  \centering
  \includegraphics[width=0.9\linewidth]{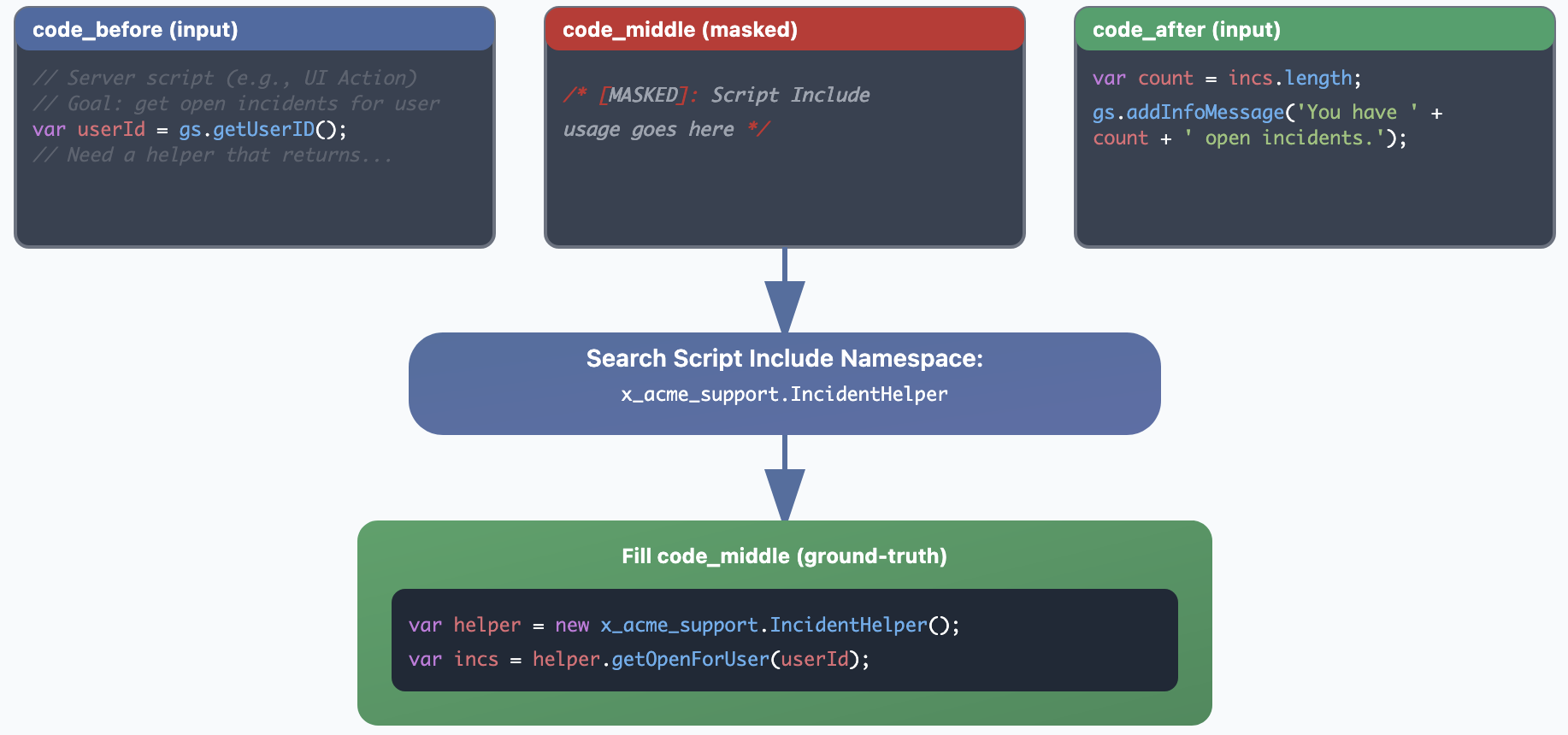}
  \caption{\textbf{Task anatomy:} highlighting how \texttt{code\_before}, and \texttt{code\_after} are used to recover the ground-truth Script Include required for \texttt{code\_middle}.}
  \label{fig:dataset_io}
\end{figure}
This setup makes the retrieval task realistic and challenging, because the model must understand the code context without direct hints. The “incomplete code” provided as the input for completion can be FIM (fill-in-the-middle) or non-FIM format depending on whether \textbf{code\_after} is available in the input.

To ensure data quality, we employed an LLM judge (Gemini 2.5 Flash) to evaluate the clarity of developer intent in each sample. The judge identified 705 samples (83\%) as having clear intent, which we use for our primary evaluation. The remaining 145 samples with ambiguous intent are excluded from our main results but analyzed separately to understand failure cases.

\subsection{Index Construction}

Our search index covers 277 distinct Script Include namespaces containing 3,337 individual APIs. The index documents vary significantly in length: full scripts range from 66 to 10,407 tokens (mean: 2,280), while corresponding JSDoc summaries are more concise, ranging from 157 to 5,368 tokens (mean: 807).

Through extensive experimentation, we found that JSDoc summaries provide superior retrieval performance compared to raw code. This is attributed to their structured nature and focused representation of API functionality. Consequently, we use JSDoc summaries for our final optimized index, which provides a cleaner signal for retrieval while maintaining comprehensive coverage of API capabilities.

\subsection{Knowledge Graph Construction}

We constructed a hierarchical Knowledge Graph from ServiceNow platform metadata to enable efficient search space pruning. The graph captures the relationship between packages, scopes, and Script Includes, allowing us to filter candidates based on contextual relevance before expensive retrieval operations. Appendix~\ref{sec:kg_analysis} details how this metadata-based graph helps reduce search space in our use case.
\section{Experimental Setup}
\label{sec:experimental_setup}

\subsection{Evaluation Metrics}

We evaluate our retrieval pipeline using two primary metrics:

\begin{itemize}
    \item \textbf{Top-K Accuracy:} The percentage of queries where the correct Script Include appears in the top-K retrieved results. We report results for K = @5, @10, @20, and @40.
    \item \textbf{Mean Reciprocal Rank (MRR):} The average of the reciprocal ranks of the correct Script Include across all queries, providing a more nuanced view of ranking quality.
\end{itemize}

\subsection{Baselines and Implementation}

We compare our proposed pipeline against BM25 \cite{robertson2009probabilistic} as the primary baseline, which achieved 53.02\% top-40 accuracy on our dataset. Our implementation uses the following components:

\begin{itemize}
    \item \textbf{Embedding Model:} Linq-AI-Research/Linq-Embed-Mistral (7B parameters, 32K context length)
    \item \textbf{Reranker Models:} Qwen-8B (baseline) and our optimized Qwen-0.6B models
    \item \textbf{Judge Model:} Gemini 2.5 Flash (1M context length) for intent clarity evaluation
\end{itemize}

\subsection{Pretraining Knowledge Check}
To test whether Script Include knowledge was already present in model pretraining corpora, we prompted LLMs to autocomplete our evaluation samples without any retrieval context (non-FIM, no KG, no index). The model produced the correct Script Include namespace in only 5\% of cases, indicating limited memorization/coverage and motivating retrieval for this domain.

\section{Experiments and Results}
\label{sec:experiments}

\subsection{Main Results}

We evaluate three primary retrieval methods against our BM25 baseline: (1) \textbf{Prefix Code Embed (Non-FIM)}, which uses embeddings of the code preceding the cursor; (2) \textbf{LLM Description}, which generates a natural language description of user intent; and (3) \textbf{Hypothetical Code Generation}, which generates hypothetical code completions for retrieval. For concise method prompts and working examples, see Appendix~\ref{sec:appendix_prompts} and Section~\ref{sec:appendix_examples}, respectively.

Table~\ref{tab:main_results} shows the performance of these methods on our clear-intent evaluation subset (705 samples). The Hypothetical Code Generation method consistently outperforms all other approaches, achieving 87.86\% top-40 accuracy, more than doubling the BM25 baseline performance. Table~\ref{tab:main_results_mrr} presents the Mean Reciprocal Rank (MRR) results, confirming the superior ranking quality of our approach. Appendix ~\ref{sec:ablation}, ~\ref{sec:ablation_1} and ~\ref{sec:ablation_2} have various ablation studies showing how each design choice (e.g., FIM vs. non-FIM formatting, context length, etc.) impacts accuracy in our dataset.

\begin{table}[htbp]
\centering
\caption{Top-K Accuracy of Retrieval Methods (Non-FIM)}
\label{tab:main_results}
\begin{tabular}{lcccc}
\toprule
\textbf{Method} & \textbf{@5 (\%)} & \textbf{@10 (\%)} & \textbf{@20 (\%)} & \textbf{@40 (\%)} \\
\midrule
BM25 (Baseline) & 26.69 & 34.16 & 40.28 & 53.02 \\
\midrule
Prefix Code Embed & 58.21 & 65.71 & 75.36 & 85.36 \\
LLM Description & 63.35 & 68.68 & 76.87 & 82.92 \\
Hypothetical Code Gen & \textbf{63.93} & \textbf{71.79} & \textbf{81.43} & \textbf{87.86} \\
\bottomrule
\end{tabular}
\end{table}

\begin{table}[htbp]
\centering
\begin{minipage}{0.57\linewidth}
  \centering
  \caption{Mean Reciprocal Rank (MRR)@K of Retrieval Methods (Non-FIM)}
  \label{tab:main_results_mrr}
  \begin{tabular}{lcccc}
  \toprule
  \textbf{Method} & \textbf{@5} & \textbf{@10} & \textbf{@20} & \textbf{@40} \\
  \midrule
  BM25 (Baseline) & 0.17 & 0.18 & 0.18 & 0.19 \\
  \midrule
  Prefix Code Embed & 0.43 & 0.44 & 0.45 & 0.45 \\
  LLM Description & 0.48 & 0.49 & 0.49 & 0.49 \\
  Hypothetical Code Gen & \textbf{0.51} & \textbf{0.52} & \textbf{0.52} & \textbf{0.53} \\
  \bottomrule
  \end{tabular}
\end{minipage}
\hfill
\begin{minipage}{0.37\linewidth}
  \centering
  \caption{Latency Analysis (Non-FIM)}
  \label{tab:latency_analysis}
  \begin{tabular}{lc}
  \toprule
  \textbf{Reranker} & \textbf{Latency (ms)} \\
  \midrule
  None & \textasciitilde22 \\
  4B @ Dense 40 (HF) & \textasciitilde342 \\
  4B @ Dense 40 (vLLM) & \textasciitilde89 \\
  8B @ Dense 40 (vLLM) & \textasciitilde121 \\
  \bottomrule
  \end{tabular}
\end{minipage}
\end{table}

\subsection{Latency Analysis}

Table~\ref{tab:latency_analysis} presents latency measurements for different pipeline configurations. Our optimized 0.6B reranker matches or exceeds the 8B model while maintaining significantly reduced latency.

\section{Post Training and Optimization}
\label{sec:post_training}

To bridge the performance gap between the 0.6B and 8B Qwen reranker for our use case while enabling real-time predictions, we developed a comprehensive post-training pipeline that optimizes compact reranker models to achieve performance comparable to much larger models.

The goal is to match or surpass an 8B reranker's ranking quality at much lower latency and cost. This matters in production: a 0.6B model fits tighter memory budgets, runs with higher concurrency, and reduces tail latency. Our SFT+RL results now exceed the 8B baseline while keeping the 2.5x latency gain, which makes the approach deployment-ready.

\subsection{Training Dataset}

A critical challenge in training our reranker was ensuring no training contamination from our evaluation datasets. To address this, we constructed completely fresh training datasets using previously unused Script Include namespaces.

\subsubsection{Dataset for SFT}

We constructed a comprehensive dataset for supervised fine-tuning using subsets of the CodeR-Pile dataset, focusing on JavaScript and TypeScript samples to establish a robust foundation for code understanding. This dataset provides the necessary diversity and scale for effective SFT training.

\subsubsection{Synthetic Dataset for RL}

For reinforcement learning we reused Script Include namespaces that never appeared in our training or evaluation data. First, we pulled 892 such namespaces (about 5.9K methods) that had been left out of earlier extraction runs. We then generated fresh JSDoc signatures with an LLM so every namespace had structured documentation in the index. Claude~3.7 analyzed each script and produced synthetic triplets---\texttt{code\_before}, \texttt{code\_middle}, \texttt{code\_after}---with the target usage placed in \texttt{code\_middle}.

We cleaned the pool in three passes. We removed 30 samples where the ground-truth namespace leaked into \texttt{code\_before} or \texttt{code\_after}, dropped another 10--15 samples that still mentioned the namespace nearby, and finally used fuzzy matching to cut 894 close variants. This left 285 strong examples. From these we kept 204 samples that offered enough hard negatives via the sentence-transformers \texttt{mine\_hard\_negatives} helper; the remaining 81 lacked suitable negatives, so we discarded them.

\subsection{Training Pipeline}

Our training approach addressed the challenge of training a small model on limited data without overfitting or catastrophic forgetting. The pipeline consisted of three main stages:

\subsubsection{Supervised Fine-Tuning (SFT)}
We began with supervised fine-tuning using the dataset described in Section~\ref{sec:post_training}:

\begin{itemize}
    \item \textbf{Parameter-Efficient Training:} Experiments with full fine-tuning and PEFT + LoRA revealed that LoRA adapters provided the best performance improvements for the 0.6B reranker while maintaining efficiency.
    
    \item \textbf{Balanced Training:} We ensured balanced training by randomly selecting either positive or negative samples during training, preventing class imbalance bias.
    
    \item \textbf{Loss Function:} We employed negative log-likelihood loss (\texttt{nll\_loss}) to optimize for the true document label (1 for positive, 0 for negative).
\end{itemize}

Training exclusively on synthetic samples led to rapid overfitting. The CodeR-Pile dataset alone provided better generalization and superior performance compared to mixed training approaches, indicating that combining synthetic and open-source data did not improve results.




\begin{figure}[t]
  \centering
  \includegraphics[width=1\columnwidth]{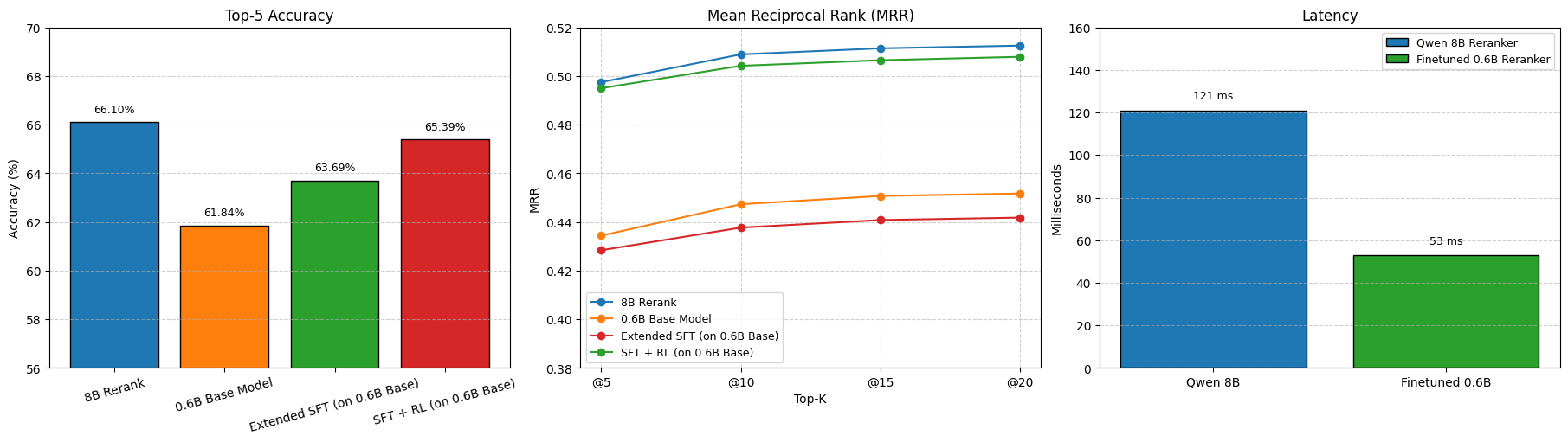}
  \caption{\textbf{Qwen 0.6B: Top-5 accuracy after finetuning, Mean Reciprocal Rank (MRR) and Inference latency}}
  \label{fig:performance_3}
\end{figure}

\subsubsection{SFT + RL}
To further improve ranking performance, we implemented a GRPO-based reinforcement learning pipeline:

\begin{itemize}
    \item \textbf{Reward Function:} Training uses a completion-aware binary reward that reads the first token produced in each GRPO rollout; matching the supervision label with ``yes'' yields $+1$, while an incorrect ``no'' response receives $-1$. Sampling eight completions per prompt injects the variance GRPO needs to shift the policy instead of collapsing toward the reference distribution. See Appendix~\ref{sec:rl_func} for the full setup and diagnostics.
    
    \item \textbf{Training Strategy:} We observed that RL training on the small synthetic dataset alone led to catastrophic forgetting. The optimal approach involved applying RL to a checkpoint from the SFT model trained on the CodeR-Pile dataset, then fine-tuning with our synthetic dataset.
    
    \item \textbf{Performance Achievement:} This two-stage approach enabled the 0.6B model to achieve performance very close to the 8B reranker on our evaluation benchmark dataset.
\end{itemize}

\subsubsection{Extended SFT}
As an alternative to RL, we experimented with extended supervised fine-tuning where we took the SFT checkpoint trained on the CodeR-Pile dataset and further fine-tuned it separately with our synthetic dataset. While this approach provided decent results, it did not show noticeable improvements over the previous SFT checkpoint.

\subsection{Training Results and Validation}

Our comprehensive evaluation demonstrates the effectiveness of the post-training pipeline. Figure~\ref{fig:performance_3} shows the complete post-training results across all stages, while Table~\ref{tab:training_results} presents the detailed performance comparison:

\begin{table}[htbp]
\centering
\caption{Training Results Comparison (FIM – Hypothetical Code Generation, Dense 40)}
\label{tab:training_results}
\begin{tabular}{lcccc}
\toprule
\textbf{Model} & \textbf{@5 (\%)} & \textbf{@10 (\%)} & \textbf{@15 (\%)} & \textbf{@20 (\%)} \\
\midrule
8B Reranker (Baseline) & 66.10 & 74.61 & 77.87 & 79.72 \\
\midrule
Qwen 0.6B Base & 61.84 & 71.35 & 75.60 & 77.30 \\
Qwen 0.6B SFT & 63.26 & 70.21 & 74.04 & 75.89 \\
Qwen 0.6B Extended SFT & 63.69 & 70.92 & 74.89 & 78.01 \\
Qwen 0.6B SFT + RL & \textbf{68.58} & \textbf{76.84} & \textbf{82.59} & \textbf{83.84} \\
\bottomrule
\end{tabular}
\end{table}

The post-training pipeline successfully bridges the performance gap between the 0.6B and 8B models, with the SFT + RL optimized 0.6B model achieving 68.58\% top-5 accuracy compared to 66.10\% for the 8B model—outperforming it by 2.48 percentage points. Detailed out-of-distribution evaluation metrics appear in Table~\ref{tab:ood_results} in Appendix~\ref{sec:ood_appendix}.

\section{Conclusion}

We present DeepCodeSeek, a comprehensive solution for real-time API retrieval in enterprise code completion scenarios. Our multi-stage retrieval pipeline achieves 87.86\% top-40 accuracy, more than doubling BM25 baseline performance while addressing the critical challenge of inferring developer intent from partial code.

Our key contributions are: (1) a novel retrieval pipeline combining knowledge graph filtering, enriched indexing with JSDoc documentation, and advanced query enhancement techniques; (2) a comprehensive post-training pipeline optimizing compact reranker models through synthetic dataset generation, supervised fine-tuning, and reinforcement learning; and (3) demonstration that our optimized 0.6B reranker now outperforms the 8B model (68.58\% vs 66.10\% top-5 accuracy) while maintaining 2.5x reduced latency.

Ablation studies show significant component contributions: knowledge graph filtering reduces search space by 59\%, enhanced indexing improves accuracy by 31 percentage points, and LLM reranking provides an additional 7 percentage point boost, enabling real-time code completion in production environments.

\subsection{Limitations and Future Work}

Our evaluation has limitations: dataset focus on Script Includes limits generalization to other code completion contexts, synthetic data generation may not capture real-world complexity, and small synthetic dataset size (204 samples) constrains training experiments.

Future work will focus on expanding data coverage through larger synthetic dataset generation and real-world data collection, refining the knowledge graph for enhanced filtering, and specializing the reranker for specific code completion tasks.

\begin{acknowledgments}
  We thank the ServiceNow AI team for their support and feedback throughout this project. We also acknowledge the contributions of the open-source community for providing the foundational models and tools that made this research possible.
\end{acknowledgments}

\section*{Declaration on Generative AI}
\noindent The author(s) have not employed any generative AI tools in the preparation of this work.

\newpage

\bibliography{egbib}

\newpage
\appendix

\section{Ablation Studies}
\label{sec:ablation}

We conduct comprehensive ablation studies to understand the contribution of each pipeline component.

\subsection{Knowledge Graph Filtering}

Our Knowledge Graph, constructed from ServiceNow platform metadata, captures hierarchical relationships across 17,701 Script Includes. Analysis reveals that 84\% of new SI usages conform to existing patterns, enabling effective search space reduction. By prioritizing globally scoped Script Includes, we reduce candidate sets by approximately 59\% before expensive retrieval operations.

\subsection{Indexing Strategy Impact}

Table~\ref{tab:index_improvement} quantifies the impact of our enhanced indexing strategy. Grouping methods under parent namespaces and enriching with JSDoc documentation provides dramatic improvements across all metrics.

\begin{table}[htbp]
\small
\setlength{\tabcolsep}{4pt}      
\centering
\caption{Impact of Enhanced Indexing (Top-K Accuracy \%)}
\label{tab:index_improvement}

\begin{tabular}{@{}l l
                S[table-format=2.2]
                S[table-format=2.2]@{}}
\toprule
\textbf{Method} & \textbf{Index Version} & {\textbf{@5}} & {\textbf{@40}} \\
\midrule
\multirow{2}{*}{Non-FIM}
  & Baseline                       & 36.71 & 54.12 \\
  & Namespace Grouping + JSDoc     & \bfseries 58.21 & \bfseries 85.36 \\
\bottomrule
\end{tabular}
\end{table}

\subsection{Reranking Analysis}

Table~\ref{tab:reranker_comparison} shows that the LLM reranker (Gemini 2.5 Flash) provides substantial improvements over cross-encoder approaches, achieving 72.60\% top-5 accuracy compared to 65.84\% for the Qwen Reranker (8B).

\begin{table}[htbp]
\centering
\caption{Reranker Comparison (Non-FIM, Dense 20)}
\label{tab:reranker_comparison}
\begin{tabular}{lccc}
\toprule
\textbf{Reranker} & \textbf{@5 (\%)} & \textbf{@10 (\%)} & \textbf{@15 (\%)} \\
\midrule
Qwen Reranker (8B) & 65.84 & 72.95 & 74.73 \\
LLM (Gemini 2.5 Flash) & \textbf{72.60} & \textbf{74.73} & \textbf{75.80} \\
\bottomrule
\end{tabular}
\end{table}

\subsection{Code Trimming and Context Length}
\begin{figure}[h]
  \centering
  \includegraphics[width=0.5\linewidth]{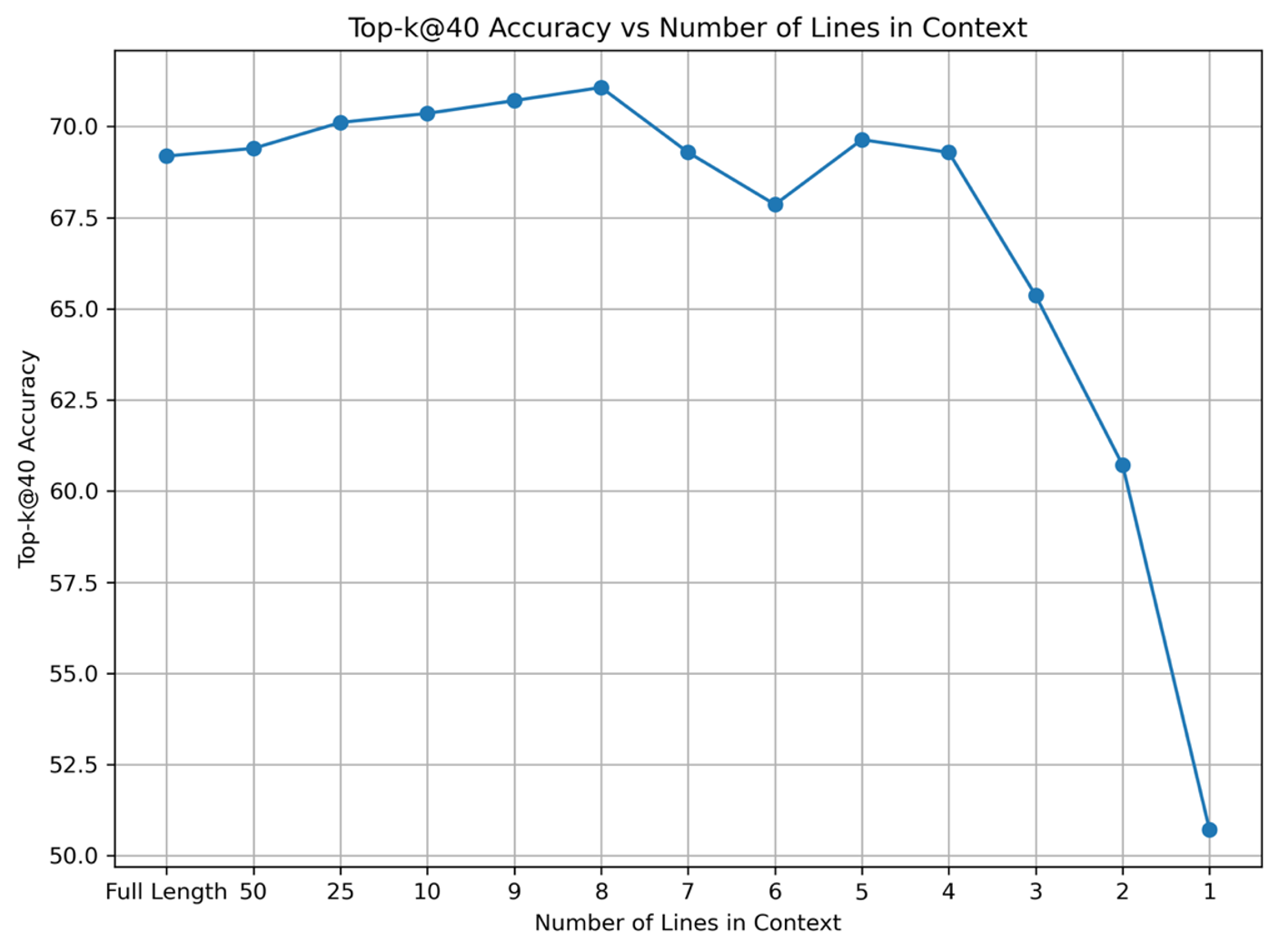}
  \caption{\textbf{A4 code trimming and context length}}
  \label{fig:code_trim}
\end{figure}

We analyzed the impact of code trimming and context length on retrieval performance. To avoid bloating the embedding model with excessive or noisy context, we experimented with various lengths of prefix code. Our experiments show that a context of 8-10 lines before the cursor yields the best performance, gaining a 1.82\% relative increase over using a larger context in our Prefix Code Embed (Non-FIM) Search. This optimal context length balances the need for sufficient information to infer developer intent while avoiding noise from distant code that may not be relevant to the current retrieval task. Note that the downstream code generation task may require more context. Figure~\ref{fig:code_trim} illustrates the relationship between context length and retrieval accuracy.

\subsection{Code Before vs. After Analysis}
\begin{figure}[h]
  \centering
  \includegraphics[width=0.9\linewidth]{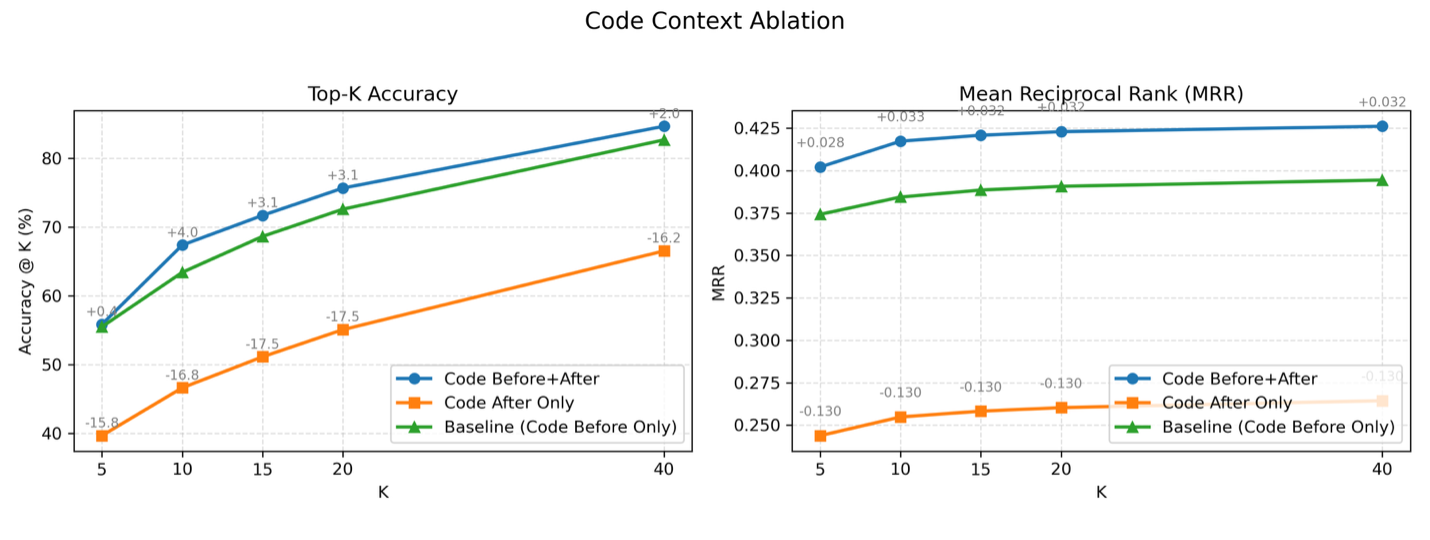}
  \caption{\textbf{Performance of Code Context Ablation }}
  \label{fig:code_context_ablation}
\end{figure}

We investigated whether using code that comes before the cursor (prefix) or after the cursor (suffix) provides better retrieval performance for user code in the middle. Our analysis revealed that the prefix code consistently outperforms suffix code for Script Include retrieval. This is likely because prefix code better captures the developer's intent and the context in which they are working, while suffix code often contains implementation details that are less useful for API retrieval. Suffix code still provides additional boost to the retrieval. Figure~\ref{fig:code_context_ablation} shows the performance comparison between using code before and after the cursor position.

\newpage

\subsection{Code Proximity and Relevance}
We conducted an ablation study to examine how the proximity of code elements affects retrieval accuracy. Our experiments revealed a critical finding: the maximum information for API retrieval is contained in the last 1-2 lines immediately preceding the API invocation. 

Our analysis compared different context trimming strategies, including using all lines before the cursor, excluding the last line, excluding the last two lines, and limiting context to ten lines with various exclusions. The results consistently showed that removing the last 1-2 lines before the API invocation leads to significant performance degradation in both Top-K accuracy and Mean Reciprocal Rank (MRR), regardless of the overall context length.

\begin{figure}[h]
  \centering
  \includegraphics[width=0.9\linewidth]{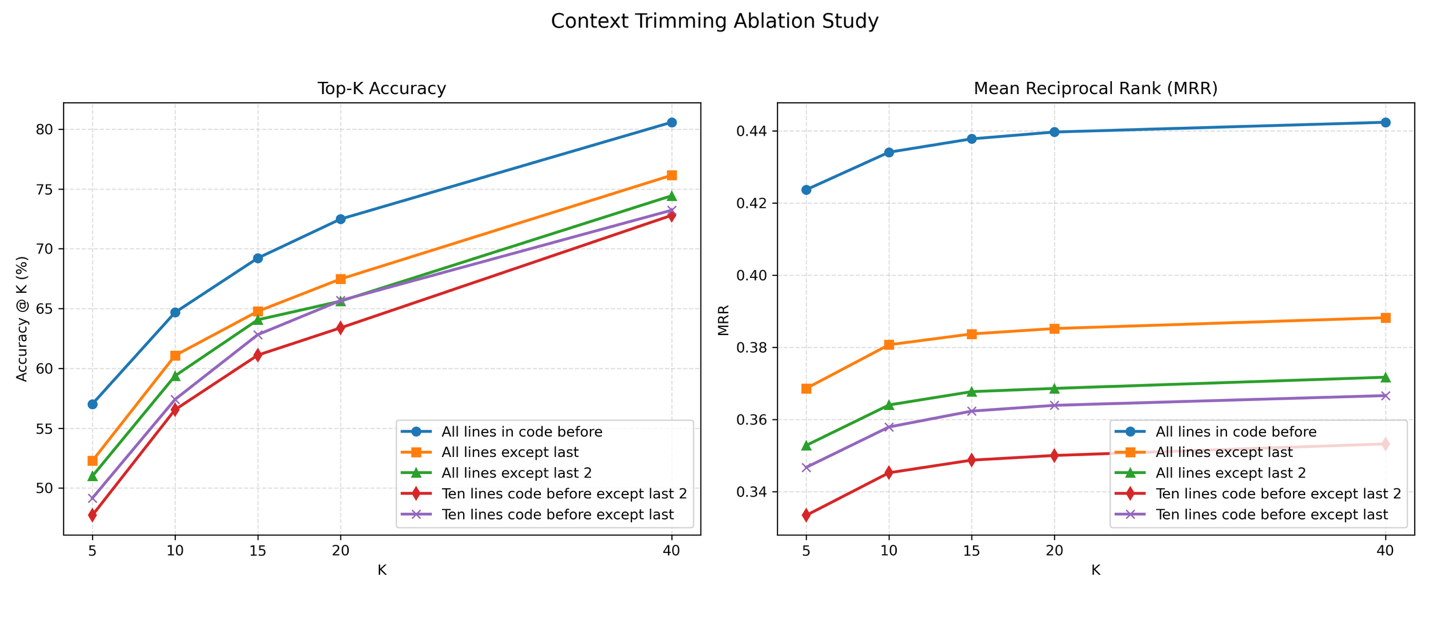}
  \caption{\textbf{Effect of Code Proximity on Performance}}
  \label{fig:code_proximity_ablation}
\end{figure}

This finding, as illustrated in ~\ref{fig:code_proximity_ablation} suggests that while broader context provides some benefit, the immediate preceding lines contain disproportionately valuable information for predicting the appropriate API. This aligns with the intuitive understanding that developers typically write code in a sequential manner, where the most recent lines provide the strongest signals about the intended functionality and API requirements.

\section{Knowledge Graph Analysis}
\label{sec:kg_analysis}

An analysis of the Script Include (SI) Knowledge Graph on our developer instance reveals several key insights. The instance contains 2,516 global SIs and 1,744 non-global SIs. By focusing the search on non-global packages and scopes, the search space is reduced by approximately ~59\%. This is a significant improvement, and the search can be narrowed even further. Approximately 97\% of non-global SIs follow a one-to-one mapping, meaning they are used in only a single package and scope. As a result, many package-scope pairs map to a single SI, often eliminating the need for a deeper search within those contexts. 

\section{Ablation Study on Model Selection}
\label{sec:ablation_1}

\begin{figure}[h]
  \centering
  \includegraphics[width=0.75\columnwidth]{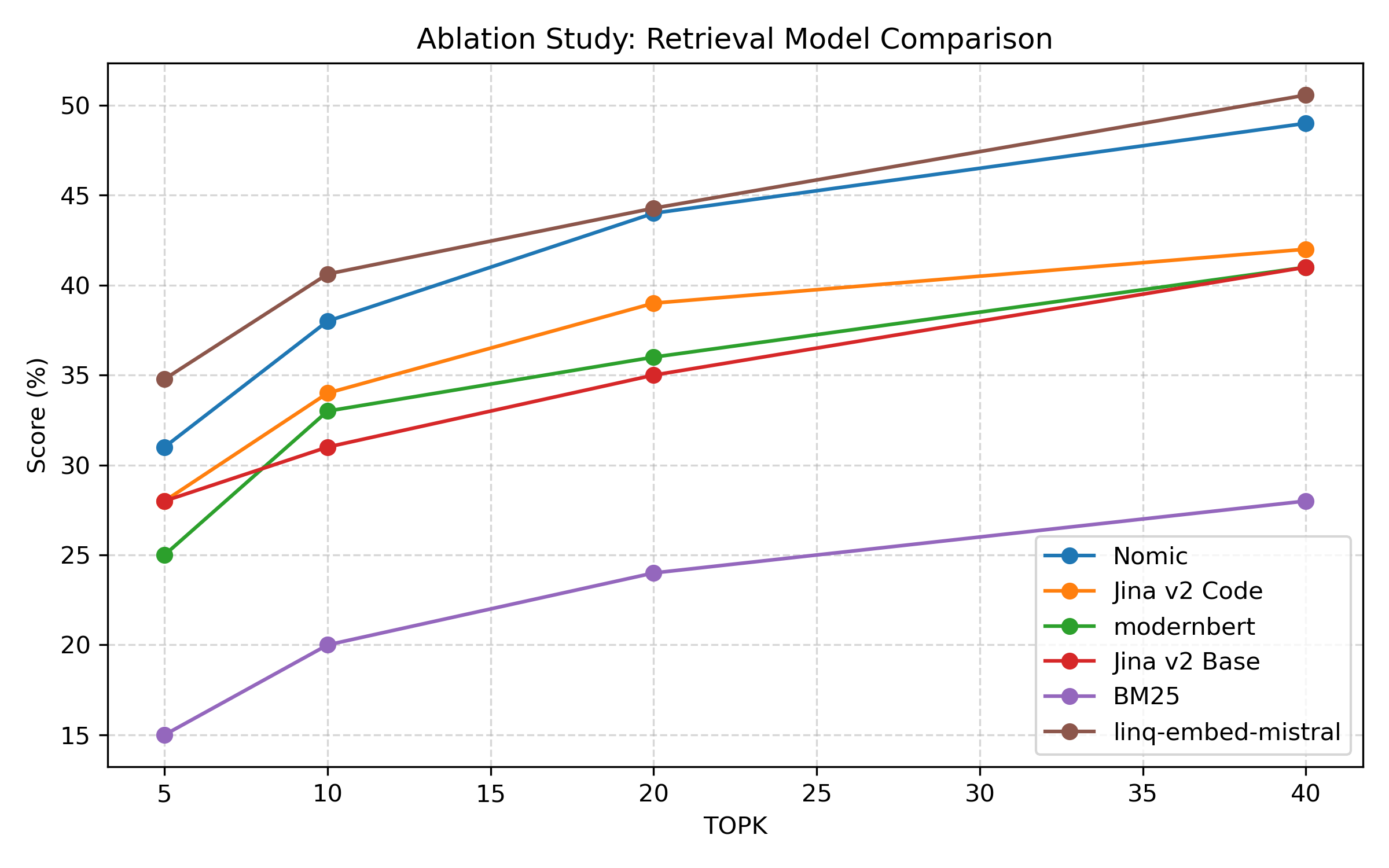}
  \caption{Ablation study comparing the retrieval performance of various models.}
  \label{fig:ablation_plot}
\end{figure}

As a preliminary step, we conducted an ablation study on an older version of our dataset to select a foundational embedding model. This initial evaluation compared several models, most notably \texttt{linq-embed-mistral} against \texttt{Jina}, which is a widely used embedding model for code retrieval. The study was performed without any advanced indexing or retrieval techniques to purely assess the baseline performance of the models. The results, shown in Figure~\ref{fig:ablation_plot}, demonstrated that \texttt{linq-embed-mistral} performed significantly better than \texttt{Jina}. Based on these preliminary findings, we chose it for all subsequent experiments in our pipeline.

\section{Code Summarization Strategy}
\label{sec:ablation_2}

Chunking the raw Script Include (SI) code proved to be ineffective and, in some cases, degraded retrieval performance. Given the availability of large-context models, we explored alternative summarization techniques. This led to the development of automated JSDoc signature generation, which creates JSDoc signatures from SI code. Using these JSDoc signatures as a concise summary of the script's functionality proved to be a more effective strategy, improving the relevance of our retrieval results.

\begin{figure}[!htb]
  \centering
  \includegraphics[width=0.7\columnwidth]{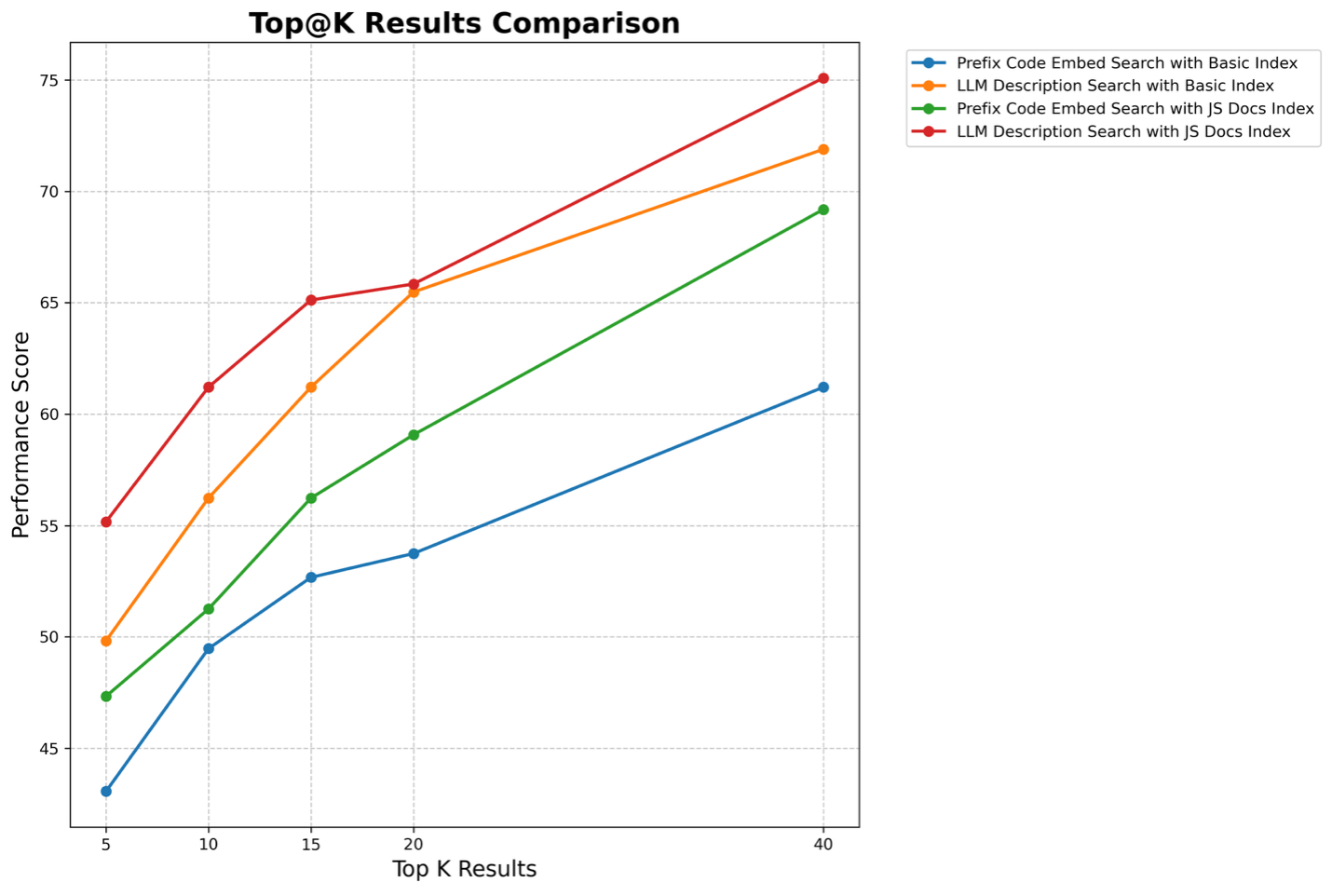}
  \caption{Performance comparison between JSDoc-based indexing and raw code descriptions across different retrieval methods. JSDoc indexing shows consistent improvements in retrieval accuracy.}
  \label{fig:jsdoc_comparison}
\end{figure}

Figure~\ref{fig:jsdoc_comparison} demonstrates the performance improvement achieved by using JSDoc documentation compared to raw code descriptions. The comparison shows that JSDoc-based indexing consistently outperforms raw code indexing across different retrieval methods, with particularly significant improvements in top-5 and top-10 accuracy metrics. This improvement is attributed to JSDoc's structured nature, which provides cleaner, more focused representations of API functionality while eliminating noise from implementation details.

\section{Training Loss Analysis}

\subsection{Supervised Fine-Tuning (SFT) Loss}

\begin{figure}[h]
  \centering
  \includegraphics[width=0.9\columnwidth]{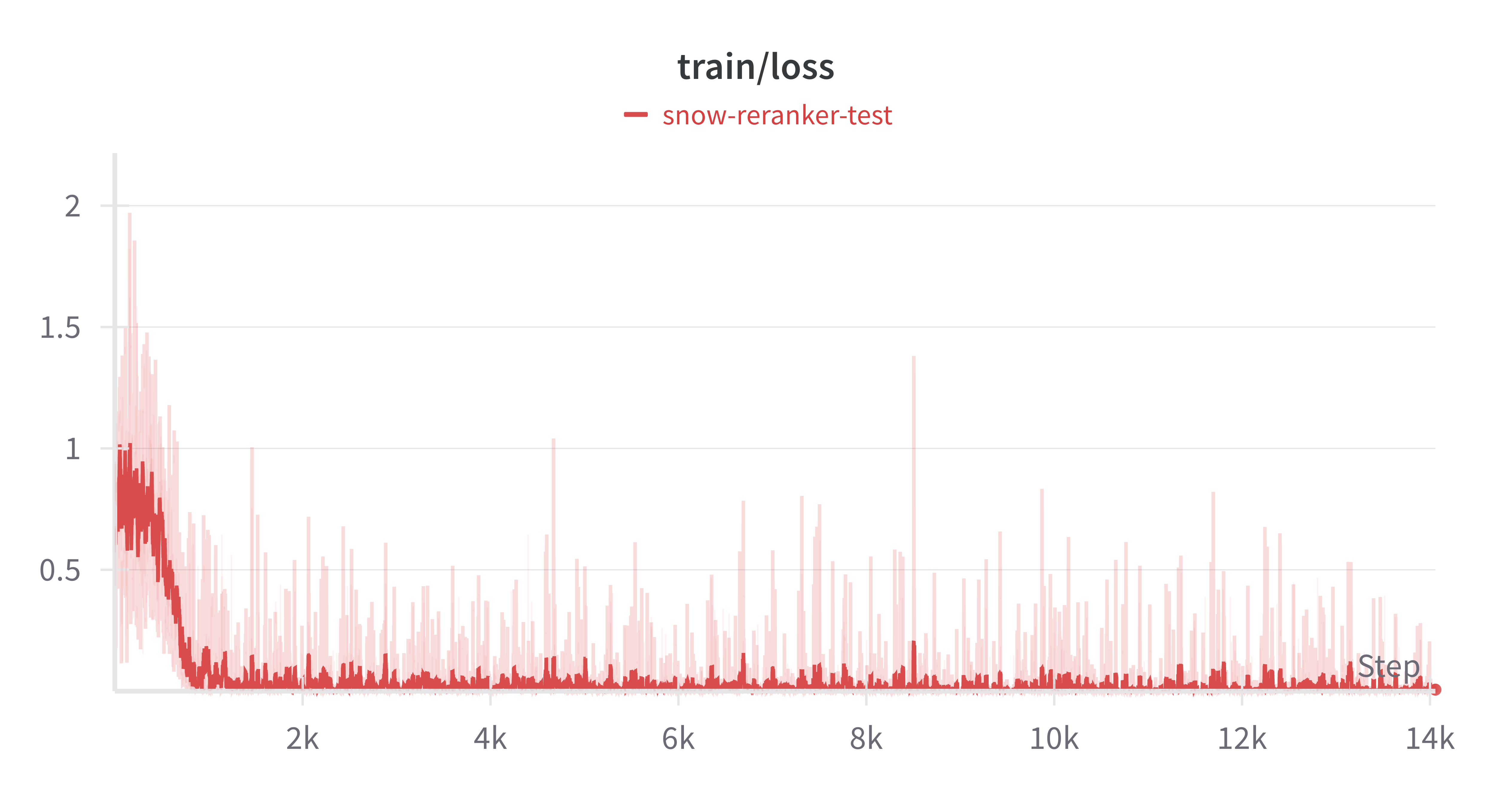}
  \caption{Training loss progression during supervised fine-tuning (SFT).}
  \label{fig:sft_train_loss}
\end{figure}

Figure~\ref{fig:sft_train_loss} shows the training loss progression during supervised fine-tuning. The model shows stable convergence with the loss decreasing steadily over epochs, indicating effective learning of the ranking task.

\begin{figure}[h]
  \centering
  \includegraphics[width=0.9\columnwidth]{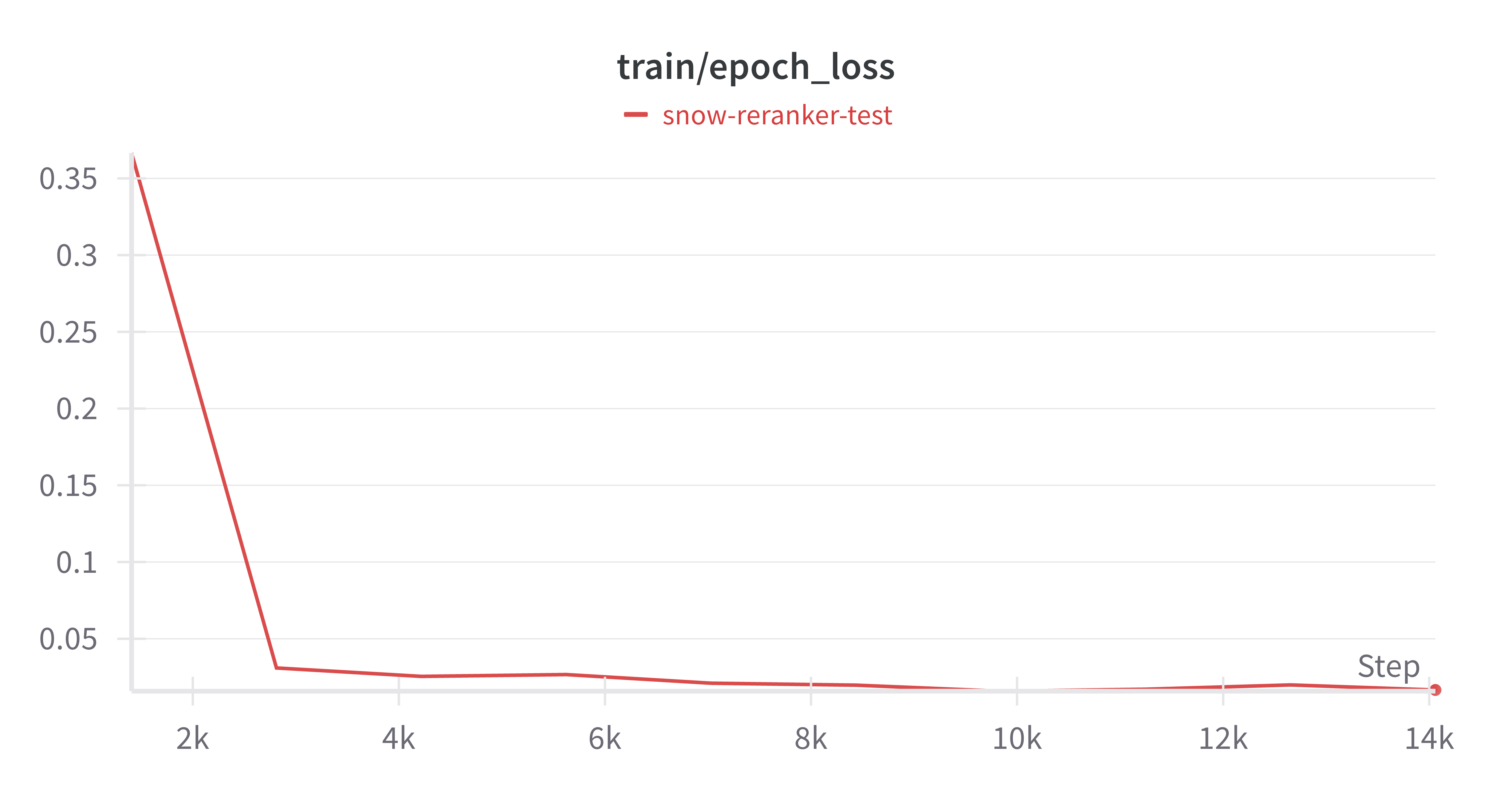}
  \caption{Epoch-wise loss during supervised fine-tuning (SFT).}
  \label{fig:sft_epoch_loss}
\end{figure}

Figure~\ref{fig:sft_epoch_loss} shows the epoch-wise loss during SFT training, providing a more granular view of the training progression across different epochs.

\subsection{Reinforcement Learning Reward Progression}

\begin{figure}[h]
  \centering
  \includegraphics[width=0.9\columnwidth]{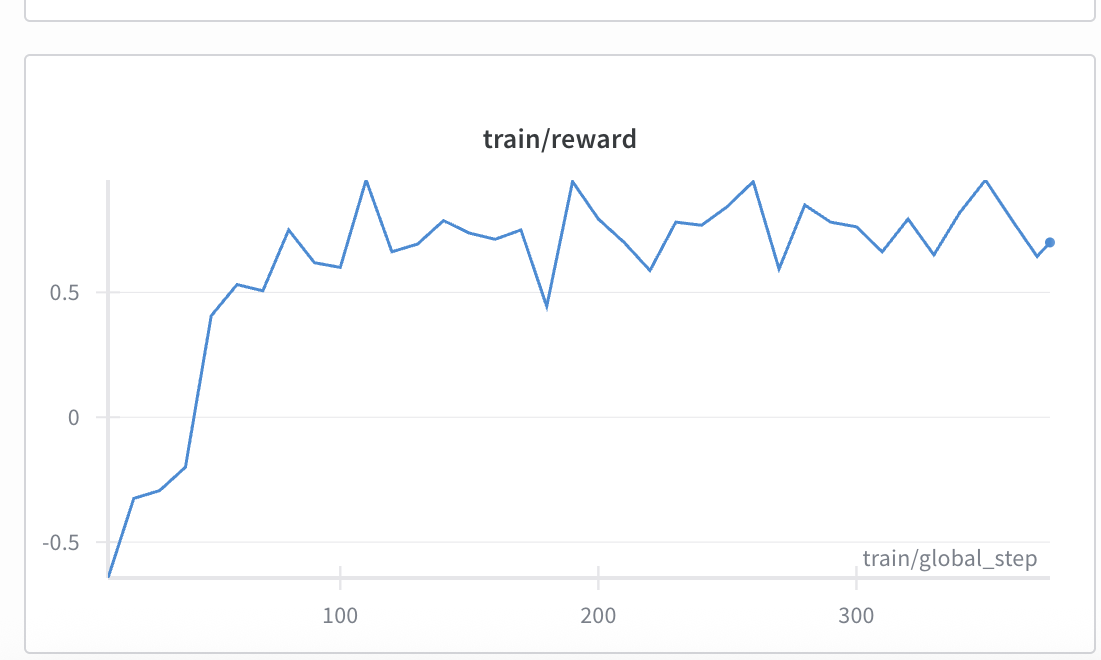}
  \caption{Reward progression during reinforcement learning training.}
  \label{fig:rl_reward}
\end{figure}

Figure~\ref{fig:rl_reward} shows the reward progression during reinforcement learning training. The completion-aware yes/no reward captures how frequently sampled responses align with the supervision label, so rising curves indicate the model is pushing more of its rollouts toward the correct decision token.

\subsection{Reinforcement Learning Training Loss}

\begin{figure}[h]
  \centering
  \includegraphics[width=0.9\columnwidth]{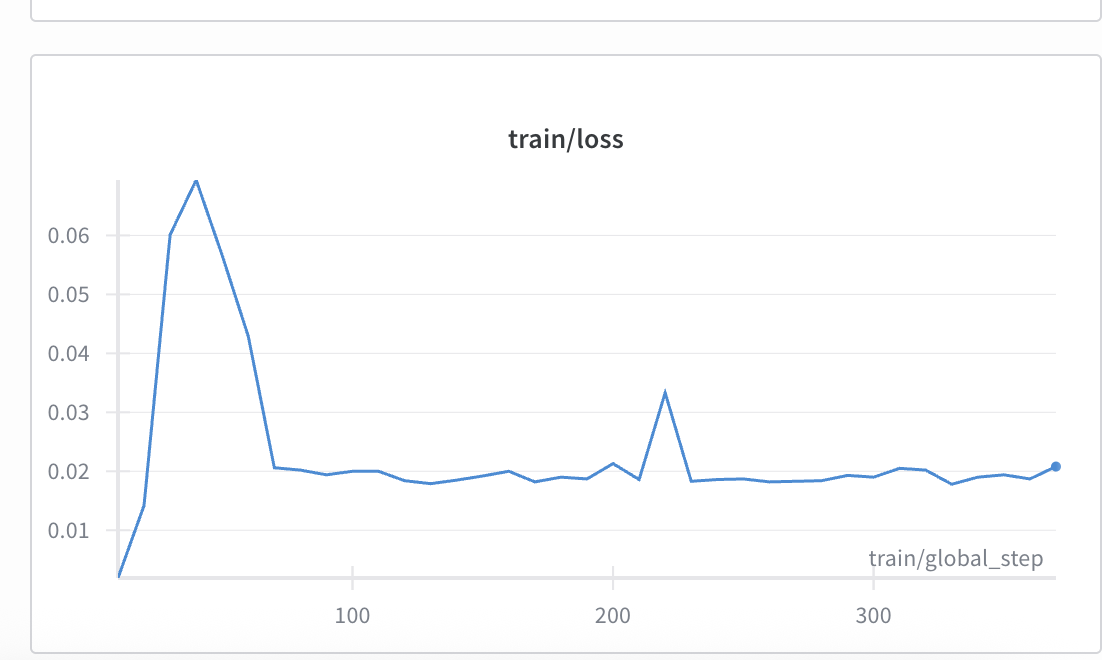}
  \caption{Training loss during reinforcement learning.}
  \label{fig:rl_train_loss}
\end{figure}

Figure~\ref{fig:rl_train_loss} shows the training loss during reinforcement learning, providing insight into the convergence behavior of the RL optimization process.

\section{Out-of-Distribution Generalization}
\label{sec:ood_appendix}

To validate that our training pipeline preserves the base model's generalization capabilities, we evaluated all trained models on an out-of-distribution dataset that none of the models had seen during training. Table~\ref{tab:ood_results} reports the results.

\begin{table}[htbp]
\centering
\caption{Out-of-Distribution Performance Validation}
\label{tab:ood_results}
\begin{tabular}{lccc}
\toprule
\textbf{Model} & \textbf{@5 (\%)} & \textbf{@10 (\%)} & \textbf{@15 (\%)} \\
\midrule
Qwen 0.6B Base & 85.50 & 89.90 & 91.30 \\
Qwen 0.6B SFT + RL & 85.30 & 88.90 & 90.70 \\
Qwen 0.6B SFT & 85.10 & 89.40 & 90.90 \\
Qwen 0.6B Extended SFT & \textbf{86.70} & \textbf{91.40} & \textbf{92.10} \\
\bottomrule
\end{tabular}
\end{table}

These results demonstrate that our trained models maintain strong generalization capabilities without suffering from catastrophic forgetting. The Qwen 0.6B models neither significantly outperform nor degrade compared to the base model performance, indicating successful specialization for our specific use case while preserving general code understanding abilities.

\section{Reward Function for Reinforcement Learning}
~\label{sec:rl_func}

Our reinforcement learning stage uses completion-aware supervision that inspects the first decision token produced by the reranker. Multiple samples per prompt provide the variance GRPO needs, while auxiliary diagnostics monitor the log-odds gap between ``yes'' and ``no'' generations.

\subsection{Completion-Based Training Reward}

The default reward passed to \texttt{GRPOTrainer} is does the following:

\begin{itemize}
    \item For every sampled completion, we read the first generated token, case-fold it, and test whether it starts with \texttt{"yes"} or \texttt{"no"}, since that is what the Qwen 0.6B reranker outputs.
    \item A correct match against the supervision label yields $+1.0$; an incorrect answer yields $-1.0$.
\end{itemize}


\subsection{Logit-Based Diagnostics}

We track the mean log-probability of answering \texttt{"yes"} on positive and negative labels, respectively.
    
These diagnostics run alongside the completion-based reward and should trend upward for positive documents once the policy improves.

\begin{figure}[h]
  \centering
  \includegraphics[width=0.7\columnwidth]{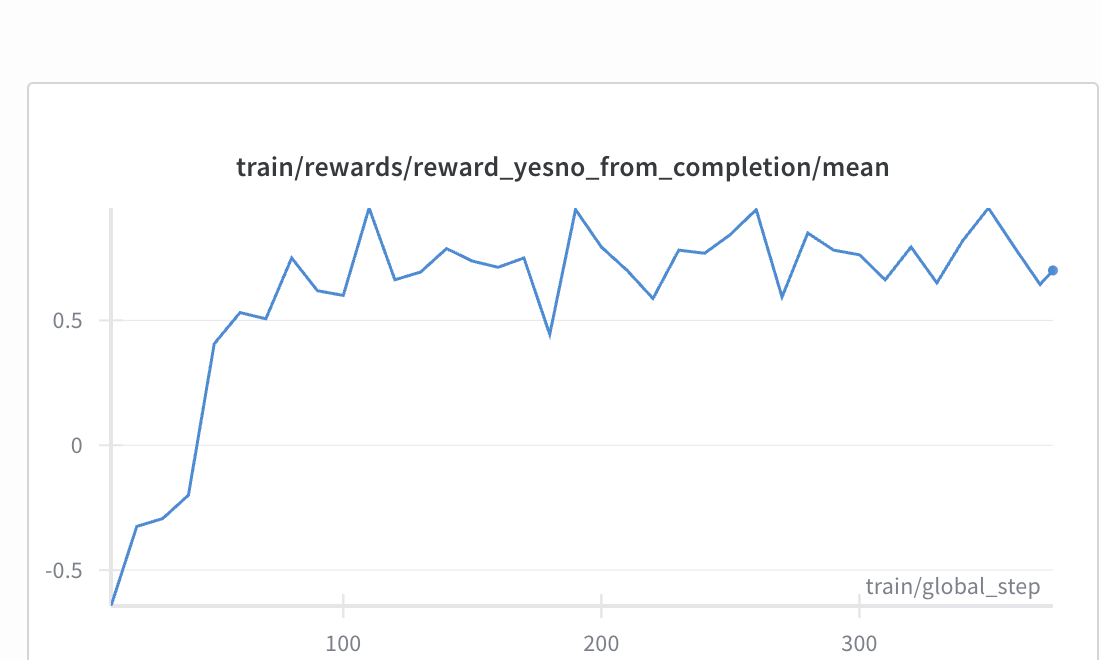}
  \caption{Log-probability of answering yes on positive labels.}
  \label{fig:rl_reward_yesno}
\end{figure}

\FloatBarrier

\section{Analysis of Query Enhancement Techniques}

Our results show that while generating a description with an LLM can help, it does not always work as well as using the code itself for the search. The main reason is that when the LLM creates a natural language summary of the user's code, it can sometimes miss important details or keywords. In contrast, the other two methods (\texttt{Prefix Code Embed (Non-FIM)} and \texttt{Hypothetical Code Generation}) use the actual code for retrieval. This provides the search model with more specific information, which likely explains why they perform better in some situations.

\section{Qualitative Examples of Retrieval Techniques}
\label{sec:appendix_examples}

To illustrate the differences between our retrieval methods, this section provides a concrete example of how each technique processes the same partial code snippet.

\subsection{Shared Context: User's Partial Code}
The following JavaScript code snippet is used as the input for all three retrieval techniques discussed below. The developer's intent is to find common elements between two arrays, a task for which the \texttt{ArrayUtil} Script Include is the correct tool.

\begin{lstlisting}[language=Java]
var prevGrp = [];
var currentGrp = [];
var commonGrp = [];
var manager;
var backupmgr;
currentGrp.push(event.parm1);
var currentGrpList = currentGrp.toString().split(",");
var grp = new GlideRecord('sys_user_group');
grp.addQuery('sys_id', event.parm2);
grp.query();
if (grp.next()) {
  manager = grp.manager;
  backupmgr = grp.u_backup_manager;
}
var grp1 = new GlideRecord('sys_user_grmember');
grp1.addQuery('group', event.parm2);
grp1.query();
while (grp1.next()) {
  prevGrp.push(grp1.user + '');
}

for (i = 0; i < currentGrpList.length; i++) {
  for (j = 0; j < prevGrp.length; j++) {
    
if (currentGrpList[i] === prevGrp[j]) {
      commonGrp.push(currentGrpList[i] + '');
    }

  }
}
\end{lstlisting}

\subsection{Technique 1: Prefix Code Embed (Non-FIM)}
This method uses a trimmed portion of the user's code directly as the search query.

\begin{lstlisting}[language=Java]
for (i = 0; i < currentGrpList.length; i++) {
  for (j = 0; j < prevGrp.length; j++) {
    
if (currentGrpList[i] === prevGrp[j]) {
      commonGrp.push(currentGrpList[i] + '');
    }

  }
}
\end{lstlisting}

\paragraph{Results}
The model successfully retrieves the correct \texttt{ArrayUtil} API as the top result.
\begin{itemize}
    \item[1.] \textbf{ArrayUtil (Correct)}
    \item[2.] Differ
    \item[3.] LiveFeedCommon
    \item[4.] XMLDocument
    \item[5.] OCGroup
\end{itemize}

\subsection{Technique 2: LLM Description}
This method uses an LLM to generate a natural language description of the user's intent, which is then used as the search query.

\begin{lstlisting}
INTENT: Identify users who are common to a previous group membership list and a new list provided by an event parameter.
\end{lstlisting}

\paragraph{Results}
The abstraction to natural language causes the correct API to be ranked second.
\begin{itemize}
    \item[1.] GlideRecordUtil
    \item[2.] \textbf{ArrayUtil (Correct)}
    \item[3.] LiveFeedCommon
    \item[4.] OCGroup
    \item[5.] LabelUpdate
\end{itemize}

\subsection{Technique 3: Hypothetical Code Generation}
This method uses an LLM to generate a hypothetical completion for the user's code. The original code context combined with this hypothetical code forms the search query.

\begin{lstlisting}[language=Java]
for (i = 0; i < currentGrpList.length; i++) {
  for (j = 0; j < prevGrp.length; j++) {
    
if (currentGrpList[i] === prevGrp[j]) {
      commonGrp.push(currentGrpList[i] + '');
    }

  }
}

/**
 * @description Finds elements that are in both currentGrpList and prevGrp arrays and adds them to the commonGrp array.
 * Completes the nested loop comparison to identify common group members between previous and current groups.
 */
if (currentGrpList[i] === prevGrp[j]) {
      commonGrp.push(currentGrpList[i] + '');
    }
\end{lstlisting}

\paragraph{Results}
This method also retrieves the correct \texttt{ArrayUtil} API as the top result, demonstrating the effectiveness of using code-based context for our search index.
\begin{itemize}
    \item[1.] \textbf{ArrayUtil (Correct)}
    \item[2.] LiveFeedCommon
    \item[3.] Differ
    \item[4.] GlideRecordUtil
    \item[5.] IdentificationLookUpTables
\end{itemize}

\section{Prompts for AI Models}
\label{sec:appendix_prompts}

This appendix details the prompts used for the various models in our pipeline.

\subsection{Instructor-based Embedding Model}
The following prompt is used to instruct the embedding model to find relevant APIs based on JSDoc for a given code snippet.

\begin{lstlisting}[language=Python]
Instruct: Given the code, find APIs based on their JSDoc that this code might need to complete its intended purpose.
Code:
\end{lstlisting}

\subsection{Reranker Model}
The reranker model uses a prefix, suffix, and an instruction to judge whether a document meets the query requirements.

\subsubsection{Prefix}
\begin{lstlisting}
Judge whether the Document meets the requirements based on the Query and the Instruct provided. Answer "yes" or "no".
\end{lstlisting}

\subsubsection{Suffix}
\begin{lstlisting}
<|im_end|>\n<|im_start|>assistant\n<think>\n\n</think>\n\n
\end{lstlisting}

\subsubsection{Instruction}
\begin{lstlisting}[language=Python]
Using the API's JSDoc, decide whether this API is directly useful for the caller-code to complete its intended task.
\end{lstlisting}

\subsection{LLM Judge for Dataset Intent}
To judge the intent of a dataset, we use a system and user prompt pair.

\subsubsection{System Prompt}
\begin{lstlisting}
Role  : ServiceNow code-completion judge
Input : (a) partial ServiceNow JavaScript Glide code, (b) proposed Script Include namespace, (c) all the possible method descriptions for the given namespace
Task  : Decide whether the namespace is the natural, intended fit for completing the partial code snippet.
Reply : Output only "Yes." if it fits, otherwise "No." (no extra text).
\end{lstlisting}

\subsubsection{User Prompt}
The user prompt is a formatted string.
\begin{lstlisting}[language=Python]
user_prompt = (
    f"### CODE:\n{code}\n\n"
    f"### NAMESPACE:\n{namespace}\n\n"
    f"### API DESCRIPTIONS (Context):\n{api_description}\n\n"
    f"Does this namespace fit the code's intent?"
)
\end{lstlisting}
\end{document}